\documentclass[fleqn, 11pt]{article}
\usepackage{amsmath, amssymb, amsthm, graphicx}

\theoremstyle{definition}
\newtheorem{theorem}{Theorem}[section]

\newtheorem{corollary}[theorem]{Corollary}

\newtheorem{Remark}[theorem]{Remark}
\newenvironment{remark}{\begin{Remark}\rm}{\end{Remark}}

\newtheorem{Example}[theorem]{Example}
\newtheorem{Question}[theorem]{Question}

\numberwithin{equation}{section}
\title{Circular non-collision orbits for a large class of $n$-body problems. }

\author{Pieter Tibboel \\
Department of Mathematical Sciences\\
Xi'an Jiaotong-Liverpool University\\
Suzhou, China\\
Pieter.Tibboel@xjtlu.edu.cn}

\begin{document}
\maketitle
\begin{abstract}
  We prove for a large class of $n$-body problems including a subclass of quasihomogeneous $n$-body problems, the classical $n$-body problem, the $n$-body problem in spaces of negative constant Gaussian curvature and a restricted case of the $n$-body problem in spaces of positive constant curvature for the case that all masses are equal and not necessarily constant that any solution for which the point masses move on a circle of not necessarily constant size has to be either a regular polygonal homographic orbit in flat space, or a regular polygonal rotopulsator in curved space, under the constraint that the minimal distance between point masses attains its minimum in finite time. Additionally, we prove that the same holds true if we add an extra mass at the center of that circle and find an explicit formula for the mass of each point particle in terms of the radius of the circle. Finally, we prove that for each order of the masses there is at most one polygonal homographic orbit for the case that the masses need not be constant.
\end{abstract}
\maketitle

\section{Introduction}
  By $n$-body problems we mean problems where we are to determine the dynamics of a number of $n$ point masses as dictated by a system of ordinary differential equations. The main $n$-body problem we will study in this paper is the problem of finding the orbits of  point masses $q_{1}$,...,$q_{n}\in\mathbb{R}^{2}$ and respective (not necessarily constant) masses $m_{1}>0$,...,\textrm{ }$m_{n}>0$ determined by the system of differential equations
  \begin{align}\label{Equations of motion}
    \ddot{q}_{i}=\sum\limits_{j=1,\textrm{ }j\neq i}^{n}m_{j}(q_{j}-q_{i})f\left(\|q_{j}-q_{i}\|^{2}\right),
  \end{align}
  where $\|\cdot\|$ is the Euclidean norm, $f:\mathbb{R}_{>0}\rightarrow\mathbb{R}_{>0}$ is a positive-valued scalar function and $\sqrt{x}f(x)$ is a decreasing function. The study of $n$-body problems of this type has applications to, for example, atomic physics, celestial mechanics, chemistry, crystallography,
  differential equations and dynamical systems (see for example \cite{AbrahamMarsden}--\cite{D0}, 
  \cite{DPS4}, 
  \cite{J}--\cite{PSSY} and the references therein and \cite{MP} and the references therein for the case that the masses are not constant).
  A second $n$-body problem we will investigate is the $n$-body problem in spaces of constant Gaussian curvature, or curved $n$-body problem for short and is defined as follows:
    Let $\sigma=\pm 1$. The $n$-body problem in spaces of constant Gaussian curvature is the problem of finding the dynamics of point masses \begin{align*}q_{1},...,\textrm{ }q_{n}\in\mathbb{M}_{\sigma}^{2}=\{(x_{1},x_{2},x_{3})\in\mathbb{R}^{3}|x_{1}^{2}+x_{2}^{2}+\sigma x_{3}^{2}=\sigma\},\end{align*} with respective masses $m_{1}>0$,..., $m_{n}>0$, determined by the system of differential equations
  \begin{align}\label{EquationsOfMotion Curved}
   \ddot{q}_{i}=\sum\limits_{j=1,\textrm{ }j\neq i}^{n}\frac{m_{j}(q_{j}-\sigma(q_{i}\odot q_{j})q_{i})}{(\sigma -\sigma(q_{i}\odot q_{j})^{2})^{\frac{3}{2}}}-\sigma(\dot{q}_{i}\odot\dot{q}_{i})q_{i}\textrm{ }i\in\{1,...,\textrm{ }n\},
  \end{align}
  where for $x$, $y\in\mathbb{M}_{\sigma}^{3}$  the product $\cdot\odot\cdot$ is defined as
  \begin{align*}
    x\odot y=x_{1}y_{1}+x_{2}y_{2}+\sigma x_{3}y_{3}.
  \end{align*}
  The curved $n$-body problem for $n=2$ goes back as far as the 1830s, but a working model for the $n\geq 2$ case was not found until 2008 by Diacu, P\'erez-Chavela and Santoprete (see \cite{DPS1}, \cite{DPS2} and \cite{DPS3}). This breakthrough then gave rise to further results for the $n\geq 2$ case in \cite{D1}--\cite{DK} and \cite{DPo}, \cite{DT} and \cite{PS}--\cite{ZZ}. See \cite{DK}, \cite{DPS1}, \cite{DPS2} and \cite{DPS3} for a historical overview. The study of the curved $n$-body problem has applications to for example geometric mechanics, Lie groups and algebras, non-Euclidean and differential geometry and stability theory, the theory of polytopes and topology (see for example \cite{D6}) and may give information about the geometry of the universe: For example: Diacu, P\'erez-Chavela and Santoprete (see \cite{DPS1}, \cite{DPS2}) showed that the configuration of the Sun, Jupiter and the Trojan asteroids cannot exist in curved space. \newline
  In this paper, we will consider solutions to (\ref{Equations of motion}) and (\ref{EquationsOfMotion Curved}) where either all point masses lie on a circle of nonconstant size, or all but one point masses lie on a circle of nonconstant size and the remaining point mass lies at the center of that circle. Specifically, we will look at solutions
  \begin{align}\label{Solution1}
    q_{i}(t)=r(t)\begin{pmatrix}
      \cos{\theta_{i}(t)}\\
      \sin{\theta_{i}(t)}
    \end{pmatrix},\textrm{ }i\in\{1,...,n\}
  \end{align}
  of (\ref{Equations of motion}), where $r$ is a twice continuously differentiable, positive function and $\theta_{1}<...<\theta_{n}$ are twice continuously differentiable functions, solutions
  \begin{align}\label{Solution2}
    q_{i}(t)=r(t)\begin{pmatrix}
      \cos{\theta_{i}(t)}\\
      \sin{\theta_{i}(t)}
    \end{pmatrix},\textrm{ }i\in\{1,...,n-1\},\textrm{ }q_{n}(t)=\begin{pmatrix}
      0 \\
      0
    \end{pmatrix}
  \end{align}
  of (\ref{Equations of motion}), where $r$ is a twice continuously differentiable, positive function and $\theta_{1}<...<\theta_{n-1}$ are twice continuously differentiable functions, solutions
  \begin{align}\label{Solution3}
    q_{i}(t)=\begin{pmatrix}
      r(t)\cos{\theta_{i}(t)}\\
      r(t)\sin{\theta_{i}(t)}\\
      z(t)
    \end{pmatrix},\textrm{ }i\in\{1,...,n\}
  \end{align}
  of (\ref{EquationsOfMotion Curved}), where $r$ is a twice continuously differentiable, positive function and $\theta_{1}<...<\theta_{n}$, $z(t)$ are twice continuously differentiable functions and solutions
  \begin{align}\label{Solution4}
    q_{i}(t)=\begin{pmatrix}
      r(t)\cos{\theta_{i}(t)}\\
      r(t)\sin{\theta_{i}(t)}\\
      z(t)
    \end{pmatrix},\textrm{ }i\in\{1,...,n-1\},\textrm{ }q_{n}(t)=\begin{pmatrix}
      0\\
      0\\
      1
    \end{pmatrix}
  \end{align}
  of (\ref{EquationsOfMotion Curved}), where $r$ is a twice continuously differentiable, positive function and $\theta_{1}<...<\theta_{n-1}$, $z(t)$ are twice continuously differentiable functions.
  Additionally, if in (\ref{Solution1}), or (\ref{Solution2}) we have that $\theta_{i}=\phi+\alpha_{i}$ for all $i\in\{1,...,N\}$, where $\phi$ is a twice, continuously differentiable function, $\alpha_{i}\in[0,2\pi)$ is constant, $N=n$ for a solution as described in (\ref{Solution1}) and $N=n-1$ for a solution as described in (\ref{Solution2}), then we call the configuration $q_{1},...,q_{n}$ a polygonal homographic orbit. If in (\ref{Solution3}), or (\ref{Solution4}) we have that $\theta_{i}=\phi+\alpha_{i}$ for all $i\in\{1,...,N\}$, where $\phi$ is a twice , continuously differentiable function, $\alpha_{i}\in[0,2\pi)$ is constant, $N=n$ for a solution as described in (\ref{Solution3}) and $N=n-1$ for a solution as described in (\ref{Solution4}), then we call the configuration $q_{1},...,q_{n}$ a polygonal rotopulsator (see \cite{DK}). If in addition in (\ref{Solution1}), (\ref{Solution2}), (\ref{Solution3}), or (\ref{Solution4}) we have that $r$ is constant, then we call respective solutions to (\ref{Equations of motion}), or (\ref{EquationsOfMotion Curved}) polygonal relative equilibria. It is well-known that polygonal homographic orbits, polygonal rotopulsators and polygonal relative equilibria exist, but whether other solutions for which the point masses move on a circle exist seems to be an unexplored and mathematically nontrivial problem. To shed light onto at least a partial answer to that question, introducing functions $\mu_{1}(t)=\min\limits_{1\leq j\leq n}\{\theta_{j+1}(t)-\theta_{j}(t)\}$, where $\theta_{n+1}=2\pi+\theta_{1}$, for solutions (\ref{Solution1}) and (\ref{Solution3}) and $\mu_{2}(t)=\min\limits_{1\leq j\leq n-1}\{\theta_{j+1}(t)-\theta_{j}(t)\}$, where $\theta_{n}=2\pi+\theta_{1}$, for solutions (\ref{Solution2}) and (\ref{Solution4}),
  we will prove the following results:
  \begin{theorem}\label{Main Theorem 1}
    Let $q_{1},...,q_{n}$ be a solution of (\ref{Equations of motion}) as described in (\ref{Solution1}), for which all masses are equal. If $\mu_{1}$ has a local minimum, then that solution is a polygonal homographic orbit with a regular polygon configuration.
  \end{theorem}
  \begin{corollary}\label{Main Theorem 1a}
    Let $q_{1},...,q_{n}$ be a solution of (\ref{Equations of motion}) as described in (\ref{Solution2}), for which all masses are equal. If $\mu_{2}$ has a local minimum, then that solution is a polygonal homographic orbit with a regular polygon configuration.
  \end{corollary}
  \begin{corollary}\label{Main Theorem 2}
    Let $q_{1},...,q_{n}$ be a solution of (\ref{EquationsOfMotion Curved}) as described in (\ref{Solution3}), for which all masses are equal, $\sigma=-1$, or $\sigma=1$ and $r<\frac{1}{5}\sqrt{10}$. If $\mu_{1}$ has a local minimum, then that solution is a polygonal rotopulsator with a regular polygon configuration.
  \end{corollary}
  \begin{corollary}\label{Main Theorem 2a}
    Let $q_{1},...,q_{n}$ be a solution of (\ref{EquationsOfMotion Curved}) as described in (\ref{Solution3}), for which all masses are equal, $\sigma=-1$, or  $\sigma=1$ and $r<\frac{1}{5}\sqrt{10}$. If $\mu_{2}$ has a local minimum, then that solution is a polygonal rotopulsator with a regular polygon configuration.
  \end{corollary}
  \begin{corollary}\label{Main Theorem 3}
    Let $q_{1},...,q_{n}$ be a polygonal homographic orbit solution of (\ref{Equations of motion}) for the case that all masses are equal. Then the configuration of the point masses is a regular polygon.
  \end{corollary}
  \begin{corollary}\label{Main Theorem 4}
    Let $q_{1},...,q_{n}$ be a polygonal homographic orbit solution of (\ref{Equations of motion}) for the case that all masses of the $q_{i}$, $i\in\{1,...,n\}$, that lie on a circle are equal to a function $m$. Then there exist constants $a$, $b\in\mathbb{R}$ such that
    \begin{align*}
      m=\frac{a^{2}-r''r^{3}}{r^{4}\sum\limits_{j=1}^{n-1}(1-\cos{\frac{2\pi j}{n}})f(2r^{2}(1-\cos{\frac{2\pi j}{n}}))}
    \end{align*}
    and
    \begin{align*}
      \phi'=\frac{a}{r^{2}}
    \end{align*}
    if the solution is as described in (\ref{Solution1}) and
    \begin{align*}
      m=\frac{b^{2}-(r''+rMf(r))r^{3}}{r^{4}\sum\limits_{j=1}^{n-2}(1-\cos{\frac{2\pi j}{n-1}})f(2r^{2}(1-\cos{\frac{2\pi j}{n-1}}))}
    \end{align*}
    and
    \begin{align*}
      \phi'=\frac{b}{r^{2}}
    \end{align*}
    if the solution is as described in (\ref{Solution2}), where $M$ is the mass of the point mass that does not lie on the circle.
  \end{corollary}
  \begin{corollary}\label{Main Theorem 5}
    If $f$ is continuously differentiable, then for each order of the masses there exists at most one polygonal homographic orbit solution of (\ref{Equations of motion}).
  \end{corollary}
  \begin{remark}
    The condition that $\mu_{1}$, or $\mu_{2}$ has a local minimum is not necessarily true for any solution $q_{1}$,...,$q_{n}$ of (\ref{Equations of motion}), or (\ref{EquationsOfMotion Curved}) where $n$, or $n-1$ point masses lie on the same circle. Proving the existence, or nonexistence, of solutions that are neither homographic orbits, nor rotopulsators where the point masses lie on a circle and $\mu_{1}$, or $\mu_{2}$ does not have a local minimum is a nontrivial matter that should be explored in future research.
  \end{remark}
  \begin{remark}
    Corollary~\ref{Main Theorem 3} was implicitly already proven in \cite{T5} for the case that $f$ in (\ref{Equations of motion}) is continuously differentiable. However, Corollary~\ref{Main Theorem 3} shows that the same result holds true if the only conditions on $f$ are that $f$ is a positive function and $x^{\frac{1}{2}}f(x)$ is a decreasing function.
  \end{remark}
  \begin{remark}
    Because of the general nature of the $n$-body problem used in Theorem~\ref{Main Theorem 1}, Corollary~\ref{Main Theorem 1a} and Corollary~\ref{Main Theorem 3}, these results may help solve Problem 12 of the list by Albouy, Cabral and Santos (see \cite{AlbouyCabralSantos}). Not only that: For the equal masses case, if the homographic orbit is a relative equilibrium, the proof of Corollary~\ref{Main Theorem 3} (or rather the proof of Theorem~\ref{Main Theorem 1}) essentially becomes a one-line proof, as $r'$ and $\theta_{i}''$, $i\in\{1,...,n\}$ are zero for polygonal relative equilibrium solutions and (\ref{Inequality 2}) reduces to $0\geq 0$, with the inequality strict if and only if the configuration is not a regular polygon.
  \end{remark}
  \begin{remark}
    Note that by Corollary~\ref{Main Theorem 4}, if $r$ is the radius of the circle on which the point masses lie, it is possible to create polygonal homographic solutions of (\ref{Equations of motion}) with almost no conditions on $r$.
  \end{remark}
  \begin{remark}
    Corollary~\ref{Main Theorem 5} was proven in \cite{T5} for the case that the homographic orbit is a relative equilibrium and all masses are constant. Corollary~\ref{Main Theorem 5} shows that the assumptions that the homographic orbit is a relative equilibrium and all masses are constant are not needed.
  \end{remark}
  We will now prove Theorem~\ref{Main Theorem 1} in section~\ref{Section proof of main theorem 1}, Corollary~\ref{Main Theorem 1a} in section~\ref{Section proof of main theorem 1a}, Corollary~\ref{Main Theorem 2} in section~\ref{Section proof of main theorem 2}, Corollary~\ref{Main Theorem 2a} in section~\ref{Section proof of main theorem 2a}, Corollary~\ref{Main Theorem 3} in section~\ref{Section proof of main theorem 3}, Corollary~\ref{Main Theorem 4} in section~\ref{Section proof of main theorem 4} and Corollary~\ref{Main Theorem 5} is section~\ref{Section proof of main theorem 5}.
  \section{Proof of Theorem~\ref{Main Theorem 1}}\label{Section proof of main theorem 1}
  Let $q_{1}$,...,$q_{n}$ be a solution of (\ref{Equations of motion}) for which all point masses lie on a sphere of radius $r$. Then we may write
  \begin{align}\label{The qi}
    q_{i}=rT(\theta_{i})\begin{pmatrix}
      1 \\
      0
    \end{pmatrix},\textrm{ }i\in\{1,...,n\},
  \end{align}
  where $r$ and the $\theta_{i}$ are scalar, differentiable functions and
  \begin{align*}
  T(x)=\begin{pmatrix}
      \cos{x} & -\sin{x} \\
      \sin{x} & \cos{x}
    \end{pmatrix}.
  \end{align*}
  Inserting (\ref{The qi}) into (\ref{Equations of motion}) and multiplying the result from the left by $T(\theta_{i})^{-1}$ gives
  \begin{align}\label{It's go-time}
    \begin{pmatrix}
      r''-r(\theta_{i}')^{2} \\
      2r'\theta_{i}'+r\theta_{i}''
    \end{pmatrix}=r\sum\limits_{j=1,\textrm{ }j\neq i}^{n}m_{j}\begin{pmatrix}
      -(1-\cos{(\theta_{j}-\theta_{i})}) \\
      \sin{(\theta_{j}-\theta_{i})}
    \end{pmatrix}f(2r^{2}(1-\cos{(\theta_{j}-\theta_{i})}))
  \end{align}
  and consequently
  \begin{align}\label{It's go-time2}
      2r'\theta_{i}'+r\theta_{i}''=r\sum\limits_{j=1,\textrm{ }j\neq i}^{n}m_{j}
      \sin{(\theta_{j}-\theta_{i})}f(2r^{2}(1-\cos{(\theta_{j}-\theta_{i})})).
  \end{align}
  If we write $\mathbb{R}_{\geq 0}=\bigcup_{k=1}^{\infty}I_{k}$, where the $I_{k}$ are disjoint intervals for which $I_{k}=[a_{k},b_{k})$ and $a_{k+1}=b_{k}$, then we can choose the $a_{k}$ and $b_{k}$ in such a way that we can write
  $\mu_{1}(t)=\theta_{2}(t)-\theta_{1}(t)$ for any $t\in I_{k}$, $k\in\mathbb{N}$, relabeling the point masses per interval $I_{k}$ if necessary. Note that $\mu_{1}$ is continuous on $\mathbb{R}_{\geq 0}$ and differentiable on the interior of every $I_{k}$.
  Note that
  \begin{align*}
    \sin{x}f(r^{2}(1-\cos{x}))=\begin{cases}
      \sqrt{1+\cos{x}}\left(\sqrt{1-\cos{x}}f(2r^{2}(1-\cos{x}))\right)\textrm{ for }x\in(0,\pi]\\
      -\sqrt{1+\cos{x}}\left(\sqrt{1-\cos{x}}f(2r^{2}(1-\cos{x}))\right)\textrm{ for }x\in(\pi,2\pi),
    \end{cases}
  \end{align*}
  so $g(x):=\sin{x}f(2r^{2}(1-\cos{x}))$ is a decreasing function on $(0,2\pi)$, as $\sqrt{y}f(y)$ for $y>0$ is a decreasing function, $1-\cos{x}$ is an increasing function and $1+\cos{x}$ is a decreasing function. Additionally, for $t\in I_{k}$, $k\in\mathbb{N}$, we have by construction that  \begin{align*}\theta_{j+1}-\theta_{j}\geq\theta_{2}-\theta_{1}\textrm{ for all }j\in\{1,...,n\},\end{align*} or equivalently
  \begin{align}\label{Crucial inequality}\theta_{j+1}-\theta_{2}\geq\theta_{j}-\theta_{1}\textrm{ for all }j\in\{1,...,n\}.\end{align}
  Also note that there are $j\in\{1,...,n\}$ for which (\ref{Crucial inequality}) is a strict inequality if and only if the configuration of the $q_{j}$ is an irregular polygon. These two observations essentially prove our theorem: As all masses are equal, write $m_{j}=m$. Then by (\ref{It's go-time2}) in combination with (\ref{Crucial inequality}) and the fact that $g(x)$ is a decreasing function on $(0,2\pi)$, we get that for $t$ in the interior of any interval $I_{k}$ that
  \begin{align}\label{Inequality 2}
    2r'\theta_{1}'+r\theta_{1}''&=r\sum\limits_{j=2}^{n}m_{j}g(\theta_{j}-\theta_{1})\geq\sum\limits_{j=2}^{n}m_{j}g(\theta_{j+1}-\theta_{2})=r\sum\limits_{j=1,\textrm{ }j\neq 2}^{n}mg(\theta_{j}-\theta_{2})\nonumber\\
    &=r\sum\limits_{j=1,\textrm{ }j\neq 2}^{n}m_{j}g(\theta_{j}-\theta_{2})=2r'\theta_{2}'+r\theta_{2}''.
  \end{align}
  So by (\ref{Inequality 2}) we now have that $2r'\theta_{2}'+r\theta_{2}''-(2r'\theta_{1}'+r\theta_{1}'')\leq 0$ for all $t\in I_{k}$, $k\in\mathbb{N}$. Multiplying both sides of this inequality with $r$ and integrating this inequality from any $s$ to a value $t$, with $s$, $t\in I_{k}$, $s<t$, $k\in\mathbb{N}$, we get that
  $r^{2}(t)(\theta_{2}'(t)-\theta_{1}'(t))\leq r^{2}(s)(\theta_{2}'(s)-\theta_{1}'(s))$, or equivalently \begin{align}\label{jackpot}r^{2}(t)\mu_{1}'(t)\leq r^{2}(s)\mu_{1}'(s).\end{align} Additionally, note that by construction $\lim\limits_{u\uparrow a_{k}}\mu_{1}'(u)\leq\lim\limits_{u\downarrow a_{k}}\mu_{1}'(u)$, which means that
  \begin{align}\label{bonus jackpot}
    \lim\limits_{u\uparrow a_{k}}r^{2}(u)\mu_{1}'(u)=r^{2}(a_{k})(\lim\limits_{u\uparrow a_{k}}\mu_{1}'(u))\leq r^{2}(a_{k})(\lim\limits_{u\downarrow a_{k}}\mu_{1}'(u))=\lim\limits_{u\downarrow a_{k}}r^{2}(u)\mu_{1}'(u).
  \end{align}
  With (\ref{jackpot}) and (\ref{bonus jackpot}) in place, we can now continue our proof: It is given that $\mu_{1}$ has a local minimum.
  If $\mu_{1}(t)$ has a local minimum for a $t=t_{0}$, $t_{0}$ in the interior of an interval $I_{k}$, then if we choose $s$ and $t$ in such a way that $s<t_{0}<t$, $s$, $t$ close enough to $t_{0}$, we get by (\ref{jackpot}) that \begin{align*}0\leq r^{2}(t)\mu_{1}'(t)\leq r^{2}(t_{0})\mu_{1}'(t_{0})\leq r^{2}(s)\mu_{1}'(s)\leq 0,\end{align*} which means that the configuration of $q_{1}$,...,$q_{n}$ is a regular polygon. If $\mu_{1}(t)$ has a local minimum for $t=a_{k}$ for a certain $k\in\mathbb{N}$, then by (\ref{jackpot}) and (\ref{bonus jackpot}) and choosing $s<a_{k}<t$, $s\in I_{k-1}$, $t\in  I_{k}$, if $k\neq 1$, $t$ and $s$ close enough to $a_{k}$, then gives that \begin{align*}0\leq r^{2}(t)\mu_{1}'(t)\leq\lim\limits_{u\uparrow a_{k}}r^{2}(u)\mu_{1}'(u))\leq \lim\limits_{u\downarrow a_{k}}r^{2}(u)\mu_{1}'(u)\leq r^{2}(s)\mu_{1}'(s)\leq 0,\end{align*} meaning that the configuration of $q_{1},...,q_{n}$ is again a regular polygon. This completes the proof.
  \section{Proof of Corollary~\ref{Main Theorem 1a}}\label{Section proof of main theorem 1a}
  Let $q_{1},...,q_{n}$ be a solution of (\ref{Equations of motion}) of the type described in (\ref{Solution2}), with $m_{n}=M$. Then
  inserting the respective $q_{j}$, $j\in\{1,...,n\}$ into (\ref{Equations of motion}), subtracting $-Mq_{i}f(\|q_{i}\|^{2})$ from both sides and then multiplying the result on both sides from the left by $T(\theta_{i})^{-1}$ gives for $i\neq n$
  \begin{align}\label{It's go-time 1a}
    \begin{pmatrix}
      r''-r(\theta_{i}')^{2} \\
      2r\theta_{i}'+r\theta_{i}''
    \end{pmatrix}+Mr\begin{pmatrix}
      1 \\ 0
    \end{pmatrix}f(r^{2})=\sum\limits_{j=1,\textrm{ }j\neq i}^{n-1}m_{j}\begin{pmatrix}
      -(1-\cos{(\theta_{j}-\theta_{i})}) \\
      \sin{(\theta_{j}-\theta_{i})}
    \end{pmatrix}f(2r^{2}(1-\cos{(\theta_{j}-\theta_{i})})).
  \end{align}
  The identity for the second coordinate in (\ref{It's go-time 1a}) is now the same as (\ref{It's go-time2}), except for the fact that now $n$ has been replaced by $n-1$. Replacing every $n$ in the proof of Theorem~\ref{Main Theorem 1} after (\ref{It's go-time2}) with $n-1$ and $\mu_{1}$ with $\mu_{2}$ thus completes the proof for this theorem.
  \section{Proof of Corollary~\ref{Main Theorem 2}}\label{Section proof of main theorem 2}
  Let $q_{1}$,...,$q_{n}$ be a solution of (\ref{EquationsOfMotion Curved}) for which all point masses lie on a sphere of radius $r$. Then we may write
  \begin{align}\label{The qic}
    q_{i}=\begin{pmatrix}
    rT(\theta_{i})\begin{pmatrix}
      1 \\
      0
    \end{pmatrix}\\
    z
    \end{pmatrix},\textrm{ }i\in\{1,...,n\},
  \end{align}
  where the $\theta_{i}$ are scalar, differentiable functions and
  \begin{align*}
    T(x)=\begin{pmatrix}
      \cos{x} & -\sin{x} \\
      \sin{x} & \cos{x}
    \end{pmatrix}.
  \end{align*}
  Inserting (\ref{The qic}) into (\ref{EquationsOfMotion Curved}) and multiplying the result for the first two coordinates from the left by $T(\theta_{i})^{-1}$ gives
  \begin{align}\label{It's go-timec}
    \begin{pmatrix}
      r''-r(\theta_{i}')^{2} \\
      2r\theta_{i}'+r\theta_{i}''
    \end{pmatrix}&=r\sum\limits_{j=1,\textrm{ }j\neq i}^{n}m_{j}\begin{pmatrix}
      -(\sigma(q_{i}\odot q_{j})-\cos{(\theta_{j}-\theta_{i})}) \\
      \sin{(\theta_{j}-\theta_{i})}
    \end{pmatrix}(\sigma -\sigma(q_{i}\odot q_{j})^{2})^{-\frac{3}{2}}\nonumber\\
    &-r\sigma(\dot{q}_{i}\odot\dot{q}_{i})\begin{pmatrix}
      1\\
      0
    \end{pmatrix},i\in\{1,...,n\}
  \end{align}
  and by the identity for the second coordinate of (\ref{It's go-timec}), using that
  \begin{align*}
    q_{i}\odot q_{j}=r^{2}\cos{(\theta_{j}-\theta_{i})}+\sigma z^{2}=-r^{2}(1-\cos{(\theta_{j}-\theta_{i})})+\sigma,
  \end{align*}
  we have that
  \begin{align}\label{It's go-time2c}
      2r\theta_{i}'+r\theta_{i}''=r\sum\limits_{j=1,\textrm{ }j\neq i}^{n}m_{j}
      \sin{(\theta_{j}-\theta_{i})}\left(2r^{2}(1-\cos{(\theta_{j}-\theta_{i})})-\sigma(r^{2}(1-\cos{(\theta_{j}-\theta_{i})}))^{2}\right)^{-\frac{3}{2}}.
  \end{align}
  Let
  \begin{align*}
    f\left(x^{2}\right)=\left(2r^{2}x^{2}-\sigma r^{4}x^{4}\right)^{-\frac{3}{2}}.
  \end{align*}
  Then
  \begin{align*}
    \frac{d}{dx}\left(xf\left(x^{2}\right)\right)&=\frac{d}{dx}\left(x^{-1}\left(2r^{2}-\sigma r^{4}x^{2}\right)^{-\frac{3}{2}}\right)\\
    &=-x^{-2}\left(2r^{2}-\sigma r^{4}x^{2}\right)^{-\frac{3}{2}}+3r^{4}\sigma\left(2r^{2}-\sigma r^{4}x^{2}\right)^{-\frac{5}{2}},
  \end{align*}
  so $x^{\frac{1}{2}}f(x)$ is a decreasing function if $\sigma=-1$. Additionally, if $\sigma=1$ and $r<\frac{1}{5}\sqrt{10}$, then $x^{\frac{1}{2}}f(x)$ is a decreasing function as well. So if $\sigma=-1$, or if $\sigma=1$, $r<\frac{1}{5}\sqrt{10}$, then (\ref{It's go-time2c}) has the required properties of (\ref{It's go-time2}) to continue as we did in the proof of Theorem~\ref{Main Theorem 1}. This completes the proof.
  \section{Proof of Corollary~\ref{Main Theorem 2a}}\label{Section proof of main theorem 2a}
  Let $q_{1}$,...,$q_{n}$ be a solution of (\ref{EquationsOfMotion Curved}) as described in (\ref{Solution4}). Then we may write for $i\neq n$ that
  \begin{align}\label{The qica}
    q_{i}=\begin{pmatrix}
    rT(\theta_{i})\begin{pmatrix}
      1 \\
      0
    \end{pmatrix}\\
    z
    \end{pmatrix},
  \end{align}
  where the $\theta_{i}$ are again scalar, differentiable functions and
  \begin{align*}
    T(x)=\begin{pmatrix}
      \cos{x} & -\sin{x} \\
      \sin{x} & \cos{x}
    \end{pmatrix}.
  \end{align*}
  Inserting (\ref{The qica}) into (\ref{EquationsOfMotion Curved}) and multiplying the result for the first two coordinates from the left by $T(\theta_{i})^{-1}$ gives for $i\neq n$ that
  \begin{align}\label{It's go-timeca}
    \begin{pmatrix}
      r''-r(\theta_{i}')^{2} \\
      2r\theta_{i}'+r\theta_{i}''
    \end{pmatrix}&=r\sum\limits_{j=1,\textrm{ }j\neq i}^{n-1}m_{j}\begin{pmatrix}
      -(\sigma(q_{i}\odot q_{j})-\cos{(\theta_{j}-\theta_{i})}) \\
      \sin{(\theta_{j}-\theta_{i})}
    \end{pmatrix}(\sigma -\sigma(q_{i}\odot q_{j})^{2})^{-\frac{3}{2}}\nonumber\\
    &+m_{n}r\begin{pmatrix}
      0-z\\
      0
    \end{pmatrix}(\sigma-\sigma z^{2})^{-\frac{3}{2}}-r\sigma(\dot{q}_{i}\odot\dot{q}_{i})\begin{pmatrix}
      1\\
      0
    \end{pmatrix}
  \end{align}
  and by the identity for the second coordinate of (\ref{It's go-timeca}), using again that
  \begin{align*}
    q_{i}\odot q_{j}=r^{2}\cos{(\theta_{j}-\theta_{i})}+\sigma z^{2}=-r^{2}(1-\cos{(\theta_{j}-\theta_{i})})+\sigma,
  \end{align*}
  we have that
  \begin{align}\label{It's go-time2ca}
      2r\theta_{i}'+r\theta_{i}''=r\sum\limits_{j=1,\textrm{ }j\neq i}^{n-1}m_{j}
      \sin{(\theta_{j}-\theta_{i})}\left(2r^{2}(1-\cos{(\theta_{j}-\theta_{i})})-\sigma(r^{2}(1-\cos{(\theta_{j}-\theta_{i})}))^{2}\right)^{-\frac{3}{2}},
  \end{align}
  which is the exact same formula as (\ref{It's go-time2c}), except for the case that $n$ has been replaced with $n-1$. The proof now holds by the same argument as in the proof of Theorem~\ref{Main Theorem 2} after (\ref{It's go-time2c}).
  \section{Proof of Corollary~\ref{Main Theorem 3}}\label{Section proof of main theorem 3}
  If $q_{1}$,...,$q_{n}$ is a polygonal homographic orbit of the type described in (\ref{Solution1}), then the $q_{i}$, $i\in\{1,...,n\}$ have exactly the properties of the $q_{i}$ in the proof of Theorem~\ref{Main Theorem 1}, with the restriction that there exist a function $\phi$ and constants $\alpha_{1},...,\alpha_{n}\in[0,2\pi)$ such that $\theta_{i}=\phi+\alpha_{i}$, $i\in\{1,...,n\}$, so in that case the result holds by exactly the same argument as in the proof of Theorem~\ref{Main Theorem 1}.
  If $q_{1}$,...,$q_{n}$ is a polygonal homographic orbit of the type described in (\ref{Solution2}), then the $q_{i}$, $i\in\{1,...,n-1\}$ have exactly the properties of the $q_{i}$ in the proof of Theorem~\ref{Main Theorem 2}, with the restriction that there exist a function $\phi$ and constants $\alpha_{1},...,\alpha_{n-1}\in[0,2\pi)$ such that $\theta_{i}=\phi+\alpha_{i}$, $i\in\{1,...,n-1\}$, so in that case the result holds by exactly the same argument as in the proof of Corollary~\ref{Main Theorem 1a}. This proves that any polygonal homographic orbit solution of (\ref{Equations of motion}) has to have the configuration of a regular polygon.
  \section{Proof of Corollary~\ref{Main Theorem 4}}\label{Section proof of main theorem 4}
  If $q_{1}$,...,$q_{n}$ is a polygonal homographic orbit of the type described in (\ref{Solution1}), then the $q_{i}$, $i\in\{1,...,n\}$ have exactly the properties of the $q_{i}$ in the proof of Theorem~\ref{Main Theorem 1}, with the restriction that there exist a function $\phi$ and constants $\alpha_{1},...,\alpha_{n}\in[0,2\pi)$ such that $\theta_{i}=\phi+\alpha_{i}$, $i\in\{1,...,n\}$. By Corollary~\ref{Main Theorem 3}, we have that the configuration of the point masses is a regular polygon, meaning that we may choose $\alpha_{i}=\frac{2\pi i}{n}$, which means that by (\ref{It's go-time}) we have that
  \begin{align}\label{Boom}
    \begin{pmatrix}
      r''-r(\phi')^{2} \\
      2r'\phi'+r\phi''
    \end{pmatrix}&=r\sum\limits_{j=1,\textrm{ }j\neq i}^{n}m\begin{pmatrix}
      -\left(1-\cos{\left(\frac{2\pi(j-i)}{n}\right)}\right) \\
      \sin{\left(\frac{2\pi(j-i)}{n}\right)}
    \end{pmatrix}f\left(2r^{2}\left(1-\cos{\left(\frac{2\pi(j-i)}{n}\right)}\right)\right)\nonumber\\
    &=rm\sum\limits_{j=1}^{n-1}\begin{pmatrix}
      -(1-\cos{\left(\frac{2\pi j}{n}\right)}) \\
      \sin{\left(\frac{2\pi j}{n}\right)}
    \end{pmatrix}f\left(2r^{2}\left(1-\cos{\left(\frac{2\pi j}{n}\right)}\right)\right).
  \end{align}
  Let $k=n-j$. As
  \begin{align*}
    &\sum\limits_{j=1}^{n-1}\sin{\left(\frac{2\pi j}{n}\right)}f\left(2r^{2}\left(1-\cos{\left(\frac{2\pi j}{n}\right)}\right)\right)\\
    &=\sum\limits_{k=1}^{n-1}\sin{\left(\frac{2\pi (n-k)}{n}\right)}f\left(2r^{2}\left(1-\cos{\left(\frac{2\pi(n-k)}{n}\right)}\right)\right)\\
    &=-\sum\limits_{k=1}^{n-1}\sin{\left(\frac{2\pi k}{n}\right)}f\left(2r^{2}\left(1-\cos{\left(\frac{2\pi k}{n}\right)}\right)\right),
  \end{align*}
  we have that
  \begin{align*}
    \sum\limits_{j=1}^{n-1}\sin{\left(\frac{2\pi j}{n}\right)}f\left(2r^{2}\left(1-\cos{\left(\frac{2\pi j}{n}\right)}\right)\right)=0,
  \end{align*}
  which means that by (\ref{Boom}) we have that
  \begin{align}\label{BoomBoom}
    \begin{pmatrix}
      r''-r(\phi')^{2} \\
      2r'\phi'+r\phi''
    \end{pmatrix}&=rm\sum\limits_{j=1}^{n-1}\begin{pmatrix}
      -\left(1-\cos{\left(\frac{2\pi j}{n}\right)}\right) \\
      0
    \end{pmatrix}f\left(2r^{2}\left(1-\cos{\left(\frac{2\pi j}{n}\right)}\right)\right),
  \end{align}
  which by the second vector components on both sides of (\ref{BoomBoom}) means that $2r'\phi'+r\phi''=0$, or $2rr'\phi'+r^{2}\phi''=0$, or equivalently that there exists a constant $a\in\mathbb{R}$ such that $r^{2}\phi'=a$, which by the first vector components on both sides of (\ref{BoomBoom}) means that
  \begin{align*}
      r''-\frac{a^{2}}{r^{3}}&=rm\sum\limits_{j=1}^{n-1}-\left(1-\cos{\left(\frac{2\pi j}{n}\right)}\right)f\left(2r^{2}\left(1-\cos{\left(\frac{2\pi j}{n}\right)}\right)\right),
  \end{align*}
  which shows that indeed
  \begin{align}\label{END}
      m=\frac{a^{2}-r''r^{3}}{r^{4}\sum\limits_{j=1}^{n-1}(1-\cos{\frac{2\pi j}{n}})f(2r^{2}(1-\cos{\frac{2\pi j}{n}}))}
    \end{align}
    if the solution is as described in (\ref{Solution1}). \\
    If $q_{1}$,...,$q_{n}$ is a polygonal homographic orbit of the type described in (\ref{Solution2}), then the $q_{i}$, $i\in\{1,...,n-1\}$ have exactly the properties of the $q_{i}$ in the proof of Theorem~\ref{Main Theorem 1a}, with the restriction that there exist a function $\phi$ and constants $\alpha_{1},...,\alpha_{n-1}\in[0,2\pi)$ such that $\theta_{i}=\phi+\alpha_{i}$, $i\in\{1,...,n-1\}$. By Corollary~\ref{Main Theorem 3}, we have that the configuration of the point masses is a regular polygon, meaning that we may choose $\alpha_{i}=\frac{2\pi i}{n-1}$, which means that by (\ref{It's go-time 1a}) we have that
    \begin{align}\label{BoomBoomBoom}
      \begin{pmatrix}
      r''-r(\theta_{i}')^{2} \\
      2r\theta_{i}'+r\theta_{i}''
    \end{pmatrix}+rM\begin{pmatrix}
      1 \\ 0
    \end{pmatrix}f(r^{2})=\sum\limits_{j=1}^{n-1}m\begin{pmatrix}
      -(1-\cos{(\frac{2\pi j}{n-1})}) \\
      \sin{(\frac{2\pi j}{n-1})}
    \end{pmatrix}f\left(2r^{2}\left(1-\cos{\left(\frac{2\pi j}{n-1}\right)}\right)\right).
    \end{align}
    Repeating the calculation that leads from (\ref{Boom}) to (\ref{END}) with (\ref{Boom}) replaced with (\ref{BoomBoomBoom}) then gives that there exists a constant $b\in\mathbb{R}$ such that
    \begin{align*}
      m=\frac{b^{2}-(r''+rMf(r))r^{3}}{r^{4}\sum\limits_{j=1}^{n-2}(1-\cos{\frac{2\pi j}{n-1}})f(2r^{2}(1-\cos{\frac{2\pi j}{n-1}}))}
    \end{align*}
    if the solution is as described in (\ref{Solution2}), where $M$ is the mass of the point mass that does not lie on the circle. This completes the proof.
    \section{Proof of Corollary~\ref{Main Theorem 5}}\label{Section proof of main theorem 5}
    If $q_{1},...,q_{n}$ is a polygonal homographic solution as in (\ref{Solution1}), then we may write that $\theta_{i}=\phi+\alpha_{i}$, $i\in\{1,...,n\}$, where $0\leq\alpha_{1}<...<\alpha_{n}<2\pi$ are constants and $\phi$ is a twice continuously differentiable function. By (\ref{It's go-time}) this means that
    \begin{align}\label{It's go-timeHOMOGRAPHIC1}
      \begin{pmatrix}
        r''-r(\phi')^{2} \\
        2r'\phi'+r\phi''
      \end{pmatrix}=r\sum\limits_{j=1,\textrm{ }j\neq i}^{n}m_{j}\begin{pmatrix}
        -(1-\cos{(\alpha_{j}-\alpha_{i})}) \\
        \sin{(\alpha_{j}-\alpha_{i})}
      \end{pmatrix}f(2r^{2}(1-\cos{(\alpha_{j}-\alpha_{i})})),\textrm{ }i\in\{1,...,n\}.
    \end{align}
    Additionally, by (\ref{Equations of motion}), using the cross product, where we interpret the $q_{i}$ as three dimensional vectors with their third components zero, we have that
    \begin{align*}
      \sum\limits_{i=1}^{n}m_{i}q_{i}\times\ddot{q}_{i}&=\sum\limits_{i=1}^{n}m_{i}q_{i}\times\sum\limits_{j=1,\textrm{ }j\neq i}^{n}m_{j}(q_{j}-q_{i})f\left(\|q_{j}-q_{i}\|^{2}\right) \\
      &=\sum\limits_{i=1}^{n}\sum\limits_{j=1,\textrm{ }j\neq i}^{n}m_{i}m_{j}q_{i}\times(q_{j}-q_{i})f\left(\|q_{j}-q_{i}\|^{2}\right) \\
      &=\sum\limits_{i=1}^{n}\sum\limits_{j=1,\textrm{ }j\neq i}^{n}m_{i}m_{j}q_{i}\times q_{j}f\left(\|q_{j}-q_{i}\|^{2}\right),
    \end{align*}
    and as
    \begin{align*}
      \sum\limits_{i=1}^{n}\sum\limits_{j=1,\textrm{ }j\neq i}^{n}m_{i}m_{j}q_{i}\times q_{j}f\left(\|q_{j}-q_{i}\|^{2}\right)&=\sum\limits_{j=1}^{n}\sum\limits_{i=1,\textrm{ }i\neq j}^{n}m_{j}m_{i}q_{j}\times q_{i}f\left(\|q_{i}-q_{j}\|^{2}\right)\\
      &=-\sum\limits_{i=1}^{n}\sum\limits_{j=1,\textrm{ }j\neq i}^{n}m_{i}m_{j}q_{i}\times q_{j}f\left(\|q_{j}-q_{i}\|^{2}\right),
    \end{align*}
    this means that
    \begin{align}\label{Cross product power}
      \sum\limits_{i=1}^{n}m_{i}q_{i}\times\ddot{q}_{i}=0.
    \end{align}
    Let \begin{align*}
    R(x)=\begin{pmatrix} \cos{x} & -\sin{x} & 0 \\
    \sin{x} & \cos{x} & 0 \\
    0 & 0 & 1\end{pmatrix}.\end{align*} Then
    \begin{align*}
      &q_{i}\times\ddot{q}_{i}\\
      &=rR(\phi+\alpha_{i})\begin{pmatrix}
        1 \\ 0 \\ 0
      \end{pmatrix}\times\left((r''-r(\phi')^{2})R(\phi+\alpha_{i})\begin{pmatrix}
        1\\ 0 \\ 0
      \end{pmatrix}+(2r'\phi'+r\phi'')R(\phi+\alpha_{i})\begin{pmatrix}
        0\\ 1\\ 0
      \end{pmatrix}\right)\\
      &=0+r(2r'\phi'+r\phi'')R(\phi+\alpha_{i})\begin{pmatrix}
        0 \\ 0 \\ 1
      \end{pmatrix},\textrm{ }i\in\{1,...,n\},
    \end{align*}
    so by (\ref{Cross product power}), we have that $r(2r'\phi'+r\phi'')=0$ and consequently that $\phi'=\frac{c}{r^{2}}$ for a constant $c\in\mathbb{R}$, which means by (\ref{It's go-timeHOMOGRAPHIC1}) that
    \begin{align}\label{It's go-timeHOMOGRAPHIC2}
      \begin{pmatrix}
        r''-\frac{c^{2}}{r^{3}} \\
        0
      \end{pmatrix}=r\sum\limits_{j=1,\textrm{ }j\neq i}^{n}m_{j}\begin{pmatrix}
        -(1-\cos{(\alpha_{j}-\alpha_{i})}) \\
        \sin{(\alpha_{j}-\alpha_{i})}
      \end{pmatrix}f(2r^{2}(1-\cos{(\alpha_{j}-\alpha_{i})})),\textrm{ }i\in\{1,...,n\}.
    \end{align}
    If we fix $t$, then $r(t)$, $r''(t)$ and the right-hand side of (\ref{It's go-timeHOMOGRAPHIC2}) are constant. Writing $A^{2}=-\frac{1}{r}\left(r''-\frac{c^{2}}{r^{3}}\right)$ (note that the right hand side of the first component of (\ref{It's go-timeHOMOGRAPHIC2}) is positive) then allows (\ref{It's go-timeHOMOGRAPHIC2}) to be written as
    \begin{align}\label{It's go-timeHOMOGRAPHIC3}
      r\begin{pmatrix}
        A^{2} \\
        0
      \end{pmatrix}=r\sum\limits_{j=1,\textrm{ }j\neq i}^{n}m_{j}\begin{pmatrix}
        (1-\cos{(\alpha_{j}-\alpha_{i})}) \\
        \sin{(\alpha_{j}-\alpha_{i})}
      \end{pmatrix}f(2r^{2}(1-\cos{(\alpha_{j}-\alpha_{i})})),\textrm{ }i\in\{1,...,n\}.
    \end{align}
    In \cite{T5} it was proven that any polygonal relative equilibrium solution of (\ref{Equations of motion}) $q_{1},...,q_{n}$, where
    \begin{align*}
      q_{i}(t)=r\begin{pmatrix}
        \cos{(At+\alpha_{i})}\\
        \sin{(At+\alpha_{i})}
      \end{pmatrix},\textrm{ }i\in\{1,...,n\},
    \end{align*}
    $r>0$, $A$, $\alpha_{1}$,...,$\alpha_{n}$ constants, has to solve a system like (\ref{It's go-timeHOMOGRAPHIC3}) (see \cite{T5}, (2.1)) and this was then used to prove that for each order of the masses at most one polygonal relative equilibrium solution exists. So for any fixed $t$, we have that (\ref{It's go-timeHOMOGRAPHIC3}) has at most one solution. So by extension, we have that for any order of the masses at most one polygonal homographic solution of (\ref{Equations of motion}) exists. \\
    If our polygonal homographic solution is of the form (\ref{Solution2}), then we repeat our argument that leads from (\ref{It's go-timeHOMOGRAPHIC1}) to (\ref{It's go-timeHOMOGRAPHIC3}) and start with (\ref{It's go-time 1a}) instead. This completes our proof.


\begin{thebibliography}{00}

  \bibitem{AbrahamMarsden} R. Abraham, J. Marsden, (1978) Foundations of Mechanics, Addison-Wesley Publishing Co. Reading, Mass.

  \bibitem{AlbouyCabralSantos} A. Albouy, H.E. Cabral, A.A. Santos, Some problems on the classical n-body problem, \textit{Celest. Mech. Dyn. Astr.} \textbf{133}, (2012) 369--375.


  \bibitem{AEKP} M. Arribas, A. Elipe, T. Kalvouridis, M. Palacios, Homographic solutions in the planar $n + 1$-body problem with quasi-homogeneous potentials, \textit{Celest. Mech. and Dyn. Astr.}, \textbf{99}(1), (2007) 1-–12.

  \bibitem{CLP} M. Corbera, J. Llibre and E. P\'erez–-Chavela, Equilibrium points and central configurations for the Lennard-Jones 2- and 3-body problems, \textit{Celestial Mechanics and Dynamical Astronomy} \textbf{89}(3), (2004) 235--266.

  \bibitem{CHG} J.M. Cors, G.R. Hall, G.E. Roberts, Uniqueness results for co-circular central configurations for power-law potentials, \textit{Physica D} \textbf{280–-281}, (2014) 44-–47.

  \bibitem{CDL} S. Craig, F. Diacu, E.A. Lacomba and E. P\'erez-Chavela, On the anisotropic Manev problem, \textit{J. Math. Phys.} \textbf{40}, (1999) 1--17.

  \bibitem{DDLMMPS} J. Delgado, F.N. Diacu, E.A. Lacomba, A. Mingarelli, V. Mioc, E. Perez, C. Stoica, The global flow of the Manev problem, \textit{J. Math. Phys.} \textbf{37}(6), (1996) 2748--2761.



  \bibitem{D0} F. Diacu, Near-collision dynamics for particle systems with quasihomogeneous potentials, \textit{J. Differential Equations} \textbf{128}, (1996) 58--77.

  \bibitem{D1} F. Diacu, On the singularities of the curved $n$-body problem, Trans. Amer. Math. Soc. \textbf{363} (2011), 2249--2264.

  \bibitem{D2} F. Diacu, Polygonal homographic orbits of the curved $n$-body problem, Trans. Amer. Math. Soc. \textbf{364} 5 (2012), 2783--2802.

  \bibitem{D3} F. Diacu, Relative equilibria in the 3-dimensional curved n-body problem, Memoirs Amer. Math. Soc. \textbf{228} 1071, (2013).

  \bibitem{D4} F. Diacu, Relative Equilibria of the Curved $N$-Body Problem, Atlantis Studies in Dynamical Systems, vol. 1, Atlantis Press, Amsterdam, 2012.

  \bibitem{D5} F. Diacu, The non-existence of centre-of-mass and linear-momentum integrals in the curved $n$-body problem,  arXiv:1202.4739, 12 p.

  \bibitem{D6} F. Diacu, The curved N-body problem: risks and rewards, Math. Intelligencer  \textbf{35} (2013), no. 3, 24--33.


  \bibitem{DK} F. Diacu, S. Kordlou, Rotopulsators of the curved N-body problem, J. Differ. Equations  \textbf{255} (2013), 2709--2750.


  \bibitem{DPS1} F. Diacu, E. P\'erez-Chavela and M. Santoprete, The n-body problem in spaces of constant curvature, arXiv:0807.1747, 54 p.

  \bibitem{DPS2} F. Diacu, E. P\'erez-Chavela and M. Santoprete, The n-body problem in spaces of constant curvature. Part I: Relative equilibria, J. Nonlinear Sci. \textbf{22} (2012), no. 2, 247--266.

  \bibitem{DPS3} F. Diacu, E. P\'erez-Chavela and M. Santoprete, The n-body problem in spaces of constant curvature. Part II: Singularities, J. Nonlinear Sci. \textbf{22} (2012), no. 2, 267--275.

  \bibitem{DPS4} F. Diacu, E. P\'erez-Chavela, M. Santoprete, Central configurations and total collisions for
        quasihomogeneous $n$-body problems, \textit{Nonlinear Analysis} \textbf{65}, (2006) 1425--1439.

  \bibitem{DPo} F. Diacu, S. Popa, All the Lagrangian relative equilibria of the curved 3-body problem have equal masses, J. Math. Phys. \textbf{55} (2014), 112701.

  \bibitem{DT} F. Diacu, B. Thorn, Rectangular orbits of the curved 4-body problem, Proc. Amer. Math. Soc. \textbf{143} (2015), 1583--1593.

  \bibitem{J} R. Jones, Central configurations with a quasihomogeneous potential function, \textit{J. Math. Phys.} \textbf{49}, 052901 (2008).

  \bibitem{Ka} T. J. Kalvouridis, A planar case of the n+1-body problem: The ring problem, \textit{Astrophys. Space Sci.} \textbf{260}:309325, 1999.












  \bibitem{M} J. C. Maxwell, On the stability of motions of Saturns rings. Macmillan and Cia., Cambridge, 1859.

  \bibitem{MS} V. Mioc, M. Stavinschi, On the Schwarzschild-type polygonal (n + 1)-body problem and on the associated restricted problem, \textit{Balt. Astron.} \textbf{7}, (1998) 637–-651.

  \bibitem{MS2} V. Mioc, M. Stavinschi, On Maxwell’s (n+1)-body problem in the manev-type field and on the associated restricted problem, \textit{Phys. Scripta} \textbf{60}, (1999) 483-–490.

  \bibitem{MP} J.C. Muzzio, A.R. Plastino, On the use and abuse of Newton's second law for variable mass problems, \textit{Celestial Mechanics and Dynamical Astronomy} \textbf{53}, (1992) 227--232.  ·

  \bibitem{P} V. Paraschiv, Central configurations and homographic solutions for the quasihomogeneous N-body problem, \textit{J. Math. Phys.} \textbf{53}, 122902 (2012).


  \bibitem{PSSY} E. P\'erez-Chavela, D. Saari, A. Susin and Z. Yan, Central configurations in the charged three body problem, \textit{Contemp. Math.} \textbf{198}, 137–-155 (1996).


  \bibitem{PS} E. P\'erez-Chavela, Juan Manuel S\'anchez-Cerritos, On the non-existence of hyperbolic polygonal relative equilibria for the negative curved $n$-body problem with equal masses, arXiv:1612.09270v1, 11p.

  \bibitem{T} P. Tibboel, Polygonal homographic orbits in spaces of constant curvature, Proc. Amer. Math. Soc. \textbf{141} (2013), 1465--1471.

  \bibitem{T2} P. Tibboel, Existence of a class of rotopulsators, J. Math. Anal. Appl.  \textbf{404} (2013), 185--191.

  \bibitem{T3} P. Tibboel, Existence of a lower bound for the distance between point masses of relative equilibria in spaces of constant curvature, J. Math. Anal. Appl. \textbf{416} (2014), 205--211.

  \bibitem{T4} P. Tibboel, Existence of a lower bound for the distance between point masses of relative equilibria for generalised quasi-homogeneous n-body problems and the curved n-body problem, J. Math. Phys. (2015) \textbf{56}, 032901.

  \bibitem{T5} P. Tibboel, Finiteness of polygonal relative equilibria for generalised quasi-homogeneous $n$-body problems and $n$-body problems in spaces of constant curvature, J. Math. Anal. Appl. \textbf{441} (2016), 183-–193.

  \bibitem{T6} P. Tibboel, Polygonal rotopulsators of the curved n-body problem, J. Math. Phys. \textbf{59} (2018), 022901.

  \bibitem{Z1} S. Zhu, Eulerian relative equilibria of the curved $3$-body problems in $\mathbf{S}^2$, Proc. Amer. Math. Soc. \textbf{142} (2014), 2837--2848.

  \bibitem{ZZ} S. Zhao, S. Zhu, Three-dimensional central configurations in $\mathbb{H}^{3}$ and $\mathbb{S}^{3}$, arXiv:1605.08730, 10p.
\end{thebibliography}
\end{document}